\documentclass[pdflatex,sn-mathphys-num]{sn-jnl}

\usepackage{graphicx}%
\usepackage{multirow}%
\usepackage{amsmath,amssymb,amsfonts}%
\usepackage{amsthm}%
\usepackage{mathrsfs}%
\usepackage[title]{appendix}%
\usepackage{xcolor}%
\usepackage{textcomp}%
\usepackage{manyfoot}%
\usepackage{booktabs}%
\usepackage{algorithm}%
\usepackage{algorithmicx}%
\usepackage{algpseudocode}%
\usepackage{listings}%

\theoremstyle{thmstyleone}%
\theoremstyle{thmstyletwo}%

\theoremstyle{thmstylethree}%

\begin{document}

\title[Kerr-induced non-Gaussianity of ultrafast bright squeezed vacuum]{Kerr-induced non-Gaussianity of ultrafast bright squeezed vacuum}


\author[1,2]{\fnm{Andrei} \sur{Rasputnyi}}

\author[3]{\fnm{Ilya} \sur{Karuseichyk}}

\author[1]{\fnm{Gerd} \sur{Leuchs}}

\author[4,5]{\fnm{Denis} \sur{Seletskiy}}

\author*[1,2,6]{\fnm{Maria} \sur{Chekhova}}\email{maria.chekhova@mpl.mpg.de}

\affil[1]{Max Planck Institute for the Science of Light, Erlangen, Germany}

\affil[2]{Department of Physics, Friedrich-Alexander-Universität Erlangen-Nürnberg (FAU), Erlangen, Germany}

\affil[3]{Laboratoire d’Optique Appliqu\'ee (LOA), CNRS, \'Ecole polytechnique, ENSTA, Institut Polytechnique de Paris, Palaiseau, France}

\affil[4]{femtoQ Laboratory, Department of Engineering Physics, Polytechnique Montréal, 2500 ch. de Polytechnique, Montréal, QC H3T 1J4, Canada}

\affil[5]{femtoQ Laboratory, Department of Physics and Astronomy, MSC07 4220, 1 University of New Mexico, Albuquerque NM, 87131-0001, United States}

\affil[6]{Faculty of Electrical and Computer Engineering, Technion—Israel Institute of
Technology, Haifa 32000, Israel}


\abstract{Non-Gaussian states of light are a critical resource for fault-tolerant quantum computing and enhanced metrology, but are typically faint and often obtained via post-selection. Here, we demonstrate the deterministic generation of a bright non-Gaussian state by introducing a Kerr nonlinearity to a macroscopic state of light called bright squeezed vacuum (BSV). To characterize the resulting state, we use a single-shot f-2f interferometer to sample its Husimi function. We observe a clear transformation from a 2D Gaussian distribution to an 'S'-shaped non-Gaussian profile, which is the direct statistical evidence of the intensity-dependent nonlinear phase. The negativity of the Wigner function, which is an intrinsic property of any pure non-Gaussian state, cannot be observed because BSV is a mixed state even under minute optical loss. However, we show that BSV can be considered as a mixture of pure  squeezed coherent states, for some of which Kerr-induced Wigner-function negativity is quite tolerant to loss. This work bridges the gap between quantum optics and ultrafast nonlinear optics, opening a path to quantum applications that require high photon flux.}



\maketitle

\section{Introduction}
\label{sec:intro}

Non-Gaussian states of light are of great interest in quantum photonic technologies \cite{Walschaers2021, Kawasaki2024, lvovsky2020}.
They are indispensable in universal quantum computing \cite{Menicucci2006}, as they enable fault-tolerant quantum algorithms and can enhance the performance of sensing and metrology techniques \cite{Tej2013}.
The best-known non-Gaussian states are the single-photon state \cite{Lvovsky2001}, the Schr\"odinger cat and superposition states \cite{Ourjoumtsev2007, Schulte2015}, and the Gottesman-Kitaev-Preskill state \cite{Larsen2025}. 
These states can be obtained through heralding, i.e., with post-selection \cite{Endo2023}.
Direct generation of a non-Gaussian state with a brightness of more than a few photons requires a very strong nonlinearity \cite{ghosh2025}. Brighter non-Gaussian states have never been obtained so far.

Here, we solve this problem by exploiting a macroscopic non-classical -- yet still Gaussian -- state of light called bright squeezed vacuum (BSV) \cite{Iskhakov2012}.
BSV originates from a strongly pumped unseeded optical parametric amplifier (OPA), which enhances the quantum fluctuations of the electromagnetic vacuum in a phase-sensitive manner.
Recently, we have generated single-mode BSV, which has the mean number of photons $N\sim10^{12}$ confined in a pulse of 25-fs \cite{Rasputnyi2024, Heimerl2025}.
Now we expose this state to Kerr nonlinearity to generate a very bright non-Gaussian state without posing any conditions for the detection.

Kerr, or third-order,  nonlinearity leads to the accumulation of a nonlinear phase, proportional to the peak intensity $I$: $\Delta \phi = n_2 I L \omega/c$, where $n_2$ is the nonlinear refractive index, $L$ the propagation length, $\omega$ the frequency, and $c$ the speed of light~ \cite{BoydNLO2008}.
This effect leads to self-phase modulation in the time domain, which is the workhorse for nonlinear spectral broadening and pulse compression \cite{Schulte2016}. 
It also plays a crucial role in soliton dynamics \cite{Turitsyn2012}. 
Importantly, Kerr nonlinearity is known to squeeze coherent states of light in the limit of weak nonlinear coupling~\cite{Bergman1991, Bergman1994,Schmitt1998,Kalinin2024}, which can be used to generate continuous-variable entanglement~\cite{Silberhorn2001} and polarization squeezing~\cite{Heersink2005,Dong2007}.
A further increase in the interaction strength results in a `banana'-shaped Wigner function exhibiting negativity \cite{Krzysztof2008, Sarlette2011, ghosh2025}.
The negativity of the Wigner function appears as a result of interference in phase space, as soon as the curvature becomes pronounced \cite{Najafabadi2023a}.
It turns out that for BSV, unlike for coherent states of light, even a weak Kerr nonlinearity can induce the negativity of the Wigner function, turning a Gaussian BSV state into a strongly non-Gaussian state \cite{Rosiek2024, Mabuchi2020}. 
This state has been predicted so far only theoretically \cite{Rosiek2024}, but has never been observed in an experiment.  

The results reported here measure the positive definite Husimi function, which is a convolution of the Wigner function with the vacuum state. In regimes of large curvature of the Husimi function, the corresponding Wigner function will have a negativity in the case of the pure state.
The main challenge addressed in this work is the low purity of the initial macroscopic Gaussian state of light. We demonstrate theoretically that BSV can be represented as a mixture of squeezed coherent states subject to random displacement, which induces quantum decoherence.
Despite this, each constituent squeezed coherent state retains a large photon number and exhibits extreme sensitivity to Kerr nonlinearity due to suppressed phase noise. 
Consequently, for sufficiently strong displacement, the outgoing states become non-Gaussian following the Kerr interaction, which results in the mixture of the non-Gaussian states.

The cornerstone of our approach lies in ultrafast optics, which allows us to reach high peak intensities.
Unlike conventional quantum optics, we are dealing with bright broadband light, which poses a challenge to the common techniques of the state characterization. 
To characterize the resulting state, we sample its Husimi function using a single-shot f-2f interferometer \cite{Ren2017}, which was recently ported for analysis of quantum light \cite{Cusson2024UP}. 
This measurement directly reveals a non-Gaussian 'S'-shaped distribution, the statistical evidence of the intensity-dependent phase shift responsible for ultimately generating Wigner negativity. 
Importantly, this bright non-Gaussian state is generated without any postselection.

\section{Experiment}\label{exp}

 \begin{figure}[h!]
     \centering
     \includegraphics[width=1.00\linewidth]{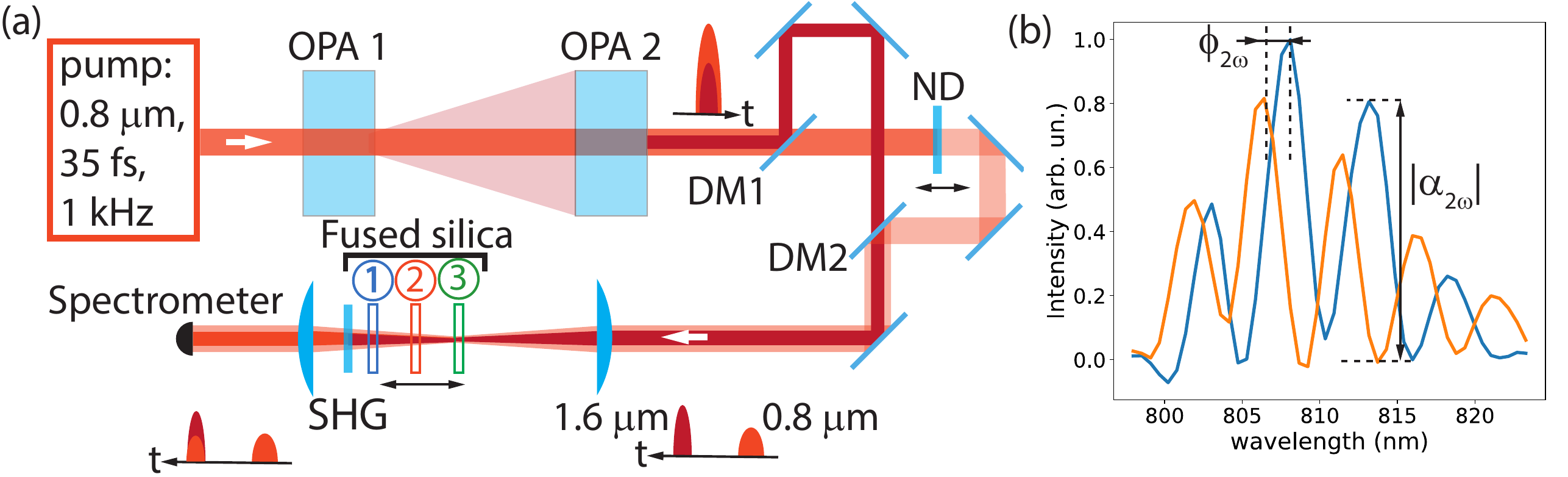}
     \caption{(a) Experimental setup. OPA 1 and OPA 2, 3-mm BBO crystals; DM1 and DM2, dichroic mirrors; ND, neutral density filter; SHG, second-harmonic generation. (b) Two typical single-shot spectra with spectral fringes (incoherent background is subtracted). Each single-shot spectrum contains information about the amplitude $|\alpha_{2\omega}|$ and relative phase $\phi_{2\omega}$ of the second-harmonic pulse of BSV.}
     \label{fig:setup}
 \end{figure}

In the experiment, we generate temporally and spatially single-mode BSV using an unseeded two-stage optical parametric amplifier (OPA 1 and OPA 2), comprising two 3~mm BBO crystals pumped by a Ti-Sa laser system (800 nm, 35 fs, 1 kHz), see Fig.~\ref{fig:setup}(a).
This allows us to obtain a 25-fs BSV centered at 1.6 $\mu$m.
Next we send both the BSV and the OPA pump to the two-color interferometer, where we separate and combine them by dichroic mirrors (DM1 and DM2) to attenuate the OPA pump by a neutral density filter (ND), rotate its polarization, to be aligned with the BSV polarization, and set the optical delay between the BSV pulse and the OPA pump pulse. 
Finally, we focus both beams using a CaF$_2$ lens into a 1.5-mm thick slab of fused silica, which imprints the Kerr phase on the BSV. 
The fused silica slab is placed at different positions after the focus, marked as (1) far from the focus, (2) at an intermediate position, and (3) at the focus, so that by moving it from position (1) to position (3) we increase the peak intensity in the slab.

Afterwards, using 0.5-mm thick periodically poled lithium niobate (PPLN), we generate the second harmonic of BSV, which interferes with the attenuated OPA pump.
As there is a delay between the second-harmonic (SH) pulse and the OPA pump pulse, we observe spectral fringes (Fig.~\ref{fig:setup}(b)).
This configuration effectively serves as a single-shot f-2f interferometer. The OPA pump pulse is phase-locked to BSV and its second harmonic (SH) because BSV is generated in a phase-sensitive manner. Therefore, we use the OPA pump pulse as a reference and estimate the amplitude $|\alpha_{2\omega}|$ and the relative phase $\phi_{2\omega}$ of each SH pulse.
The SH field scales as the square of the BSV field such that we can restore the amplitude $|\alpha_{\omega}|$ and the phase $\phi_{\omega}$ of the BSV after the Kerr interaction through the following mapping: $|\alpha_{\omega}| = \sqrt{|\alpha_{2\omega}|}$, $\phi_{\omega} = \phi_{2\omega}/2$.
By acquiring the statistics of the spectral interferogram at a fixed time delay, we can sample the Husimi function as the joint probability distribution of the amplitude and phase.

\section{'S'-shaped Husimi function}\label{sec:S}

 \begin{figure}[h!]
     \centering
     \includegraphics[width=1.00\linewidth]{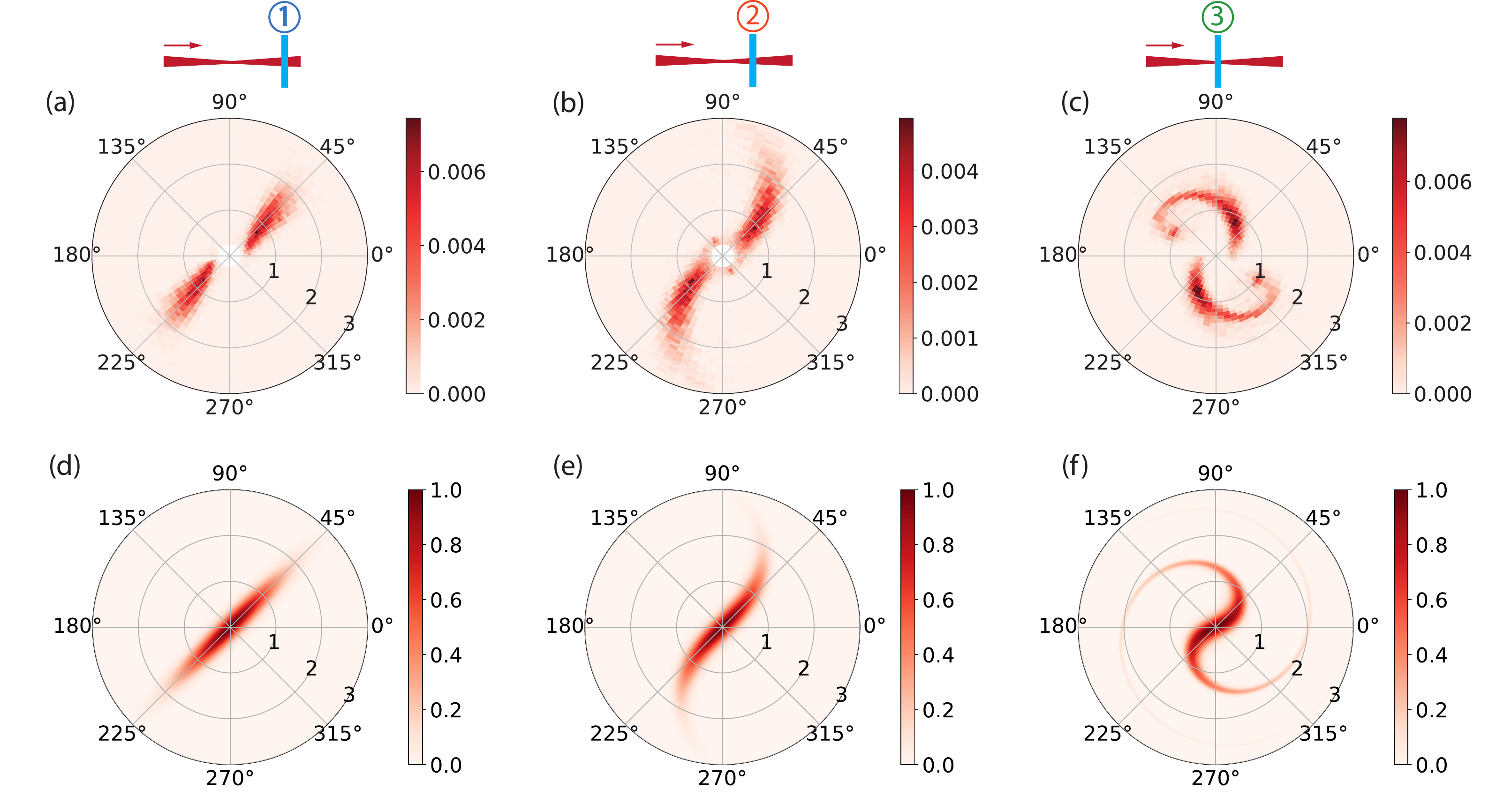}
      \caption{(a-c) Experimental and  (d-f) simulated Husimi functions for different strengths of the Kerr interaction. The radial coordinate is normalized to the mean amplitude.
      }
     \label{fig:husimi}
 \end{figure}

Figures~\ref{fig:husimi}(a-c) show the measured Husimi functions of BSV after the Kerr nonlinear interaction (Fig.~\ref{fig:setup}(a)).
When the fused silica slab is far from the focus, we observe predominantly amplitude fluctuations, typical for BSV (Fig.~\ref{fig:husimi}(a)).
Near the center of the polar plot (for small field amplitudes) the histogram shows no detectable signal because the efficiency of the second harmonic drops.
As the Kerr interaction gets stronger (position "2" of the slab), an additional nonlinear phase is observed for larger amplitudes, leading to an 'S'-shaped Husimi function (Fig.~\ref{fig:husimi}(b)).
Finally, if the sample is at the focus (position "3" of the slab), a much stronger Kerr phase is imprinted onto the original state (Fig.~\ref{fig:husimi}(c)).

To theoretically describe the large-scale structure of the measured Husimi function for a single-mode BSV under Kerr nonlinearity, it is sufficient to use classical Liouville equation (see Methods, section~\ref{sec:kerr_simm}).
In this regime, the Kerr nonlinearity reduces to a phase-space rotation at an angle proportional to the squared distance from the origin, i.e., to the peak intensity.
This amplitude-dependent shear maps the initially elongated Gaussian Husimi function of BSV into the characteristic 'S'-shaped profile shown in Fig.~\ref{fig:husimi}(d-f). 
In stark contrast, the measured distribution for the same strength of Kerr interaction (Fig.~\ref{fig:husimi}(c)) shows that the spiral tails of Husimi function are 'cut,' and the probability at large amplitudes is suppressed.
This observation indicates that at this high intensity, the desired Kerr effect is competing with other nonlinear processes, such as those related to nonlinear spectral broadening (supercontinuum generation) and non-degenerate four-wave mixing. Because these nonlinear effect are especially pronounced for high intensities, they deplete the high-photon-number components of the BSV state more than the rest.
This behavior in phase-space is consistent with the 'sharp cut-off', independently observed in the photon-number statistics for the exact same setting (see Supplementary materials, Figure S1(f), green curve).

\section{BSV as a mixture of squeezed states}\label{sec:BSV_mix}
The 'S'-shaped Husimi function is a direct evidence of the Kerr nonlinearity acting on the macroscopic state of light.
For a pure state, this nonlinear phase would lead to a Wigner-function negativity \cite{Walschaers2021, Mabuchi2020}. However, BSV is not a pure state: its purity degrades even under minute optical losses, which inevitably occur on optical elements and through additional nonlinear-optical processes accompanying optical parametric amplification.

The purity of a squeezed vacuum state can be calculated as the inverse product of its maximal and minimal quadrature uncertainties~\cite{Wenger2004}. 
From this definition, we see that losses on the order of $\sim 1 / N$, where $N$ is the BSV mean photon number, will reduce its purity by about a factor of two \cite{Leuchs2005a} (see also Methods). 
Practically, it is impossible to manage such tiny losses for  $ N\sim10^{12}$; therefore, BSV is never a pure quantum state. 
At the same time, under linear losses, the minimal variance of the BSV quadratures remains below the vacuum level, so that the state is still non-classical.

Using the fact that BSV remains a Gaussian state of light even after linear losses, although with reduced quadrature squeezing, one can show (see Methods) that after loss, BSV becomes a squeezed thermal state, which can be also seen as a mixture of squeezed coherent states $|\beta, r\rangle$,
\begin{equation}
    \hat{\rho}_{BSV} = \int d^2\beta~ \widetilde{P}(\beta) |\beta, r\rangle \langle \beta, r|,
    \label{eq:sq_th}
\end{equation}
where, for moderate losses $R$, the squeezing parameter $r$ is related to the initial squeezing parameter $r_0$ as $r\approx\frac{r_0}{2}+\frac{1}{4}\ln(1/R-1)$ and
\begin{equation}
    \widetilde{P}(\beta) = \dfrac{1}{\pi n_{th}} \exp\left[ -\dfrac{\text{Re}(\beta)^2}{n_{th} e^{2r}}-\dfrac{\text{Im}(\beta)^2}{n_{th} e^{-2r}} \right]
    \label{eq:GSF}
\end{equation} 
is the statistical weight of a squeezed coherent state in the mixture. The initial mean photon number of BSV $N$ and the amount of optical losses $R$ are mapped to  the number of thermal photons $n_{th}$, related to the optical losses, and the effective squeezing parameter $r$.

 \begin{figure}[h!]
     \centering
     \includegraphics[width=1.00\linewidth]{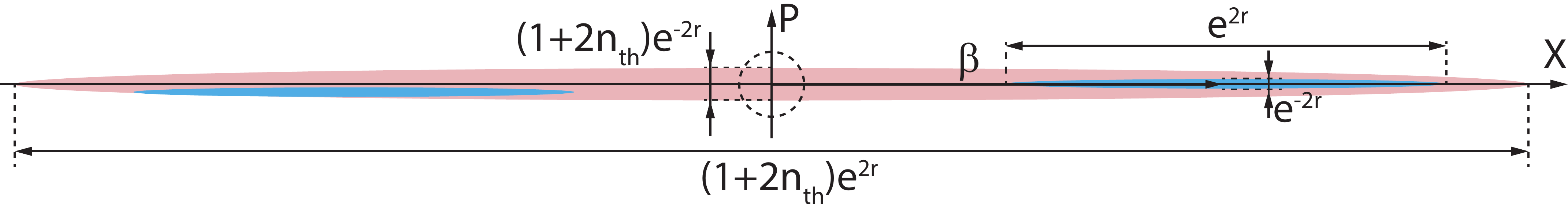}
     \caption{Schematic representation of BSV as a mixture (red ellipse) of  squeezed coherent states (blue ellipses) in phase space. The dashed circle represents the vacuum state (the shot-noise uncertainty).}
     \label{fig:husimi}
 \end{figure}

As a result, after losses BSV becomes a mixture of pure squeezed coherent states $|\beta, r\rangle$ stochastically displaced mostly in the $x$-quadrature direction (Fig. \ref{fig:husimi}), all squeezed along the same quadrature $p$, which for $\rm {Re}(\beta)\gg1$ approximately corresponds to phase squeezing. 
The effective squeezing parameter $r$ defines the area of quantum coherence in the phase space. 
Meanwhile, displacement $\beta$ acts as a stochastic variable responsible for the quantum decoherence of BSV. This stochasticity, with the probability distribution given in Eq.(\ref{eq:GSF}), washes out the negativity of the Wigner function of the non-Gaussian bright state. However, as we show in this work, those pure states in the mixture which are squeezed along the phase quadrature become strongly non-Gaussian as a result of the Kerr effect, and under reasonable loss, their Wigner functions maintain negativity even for rather high photon numbers.

\section{A phase-squeezed state and the Kerr effect}\label{sec:Kerr_sq}

\begin{figure}[h!]
     \centering
\includegraphics[width=0.66\linewidth]{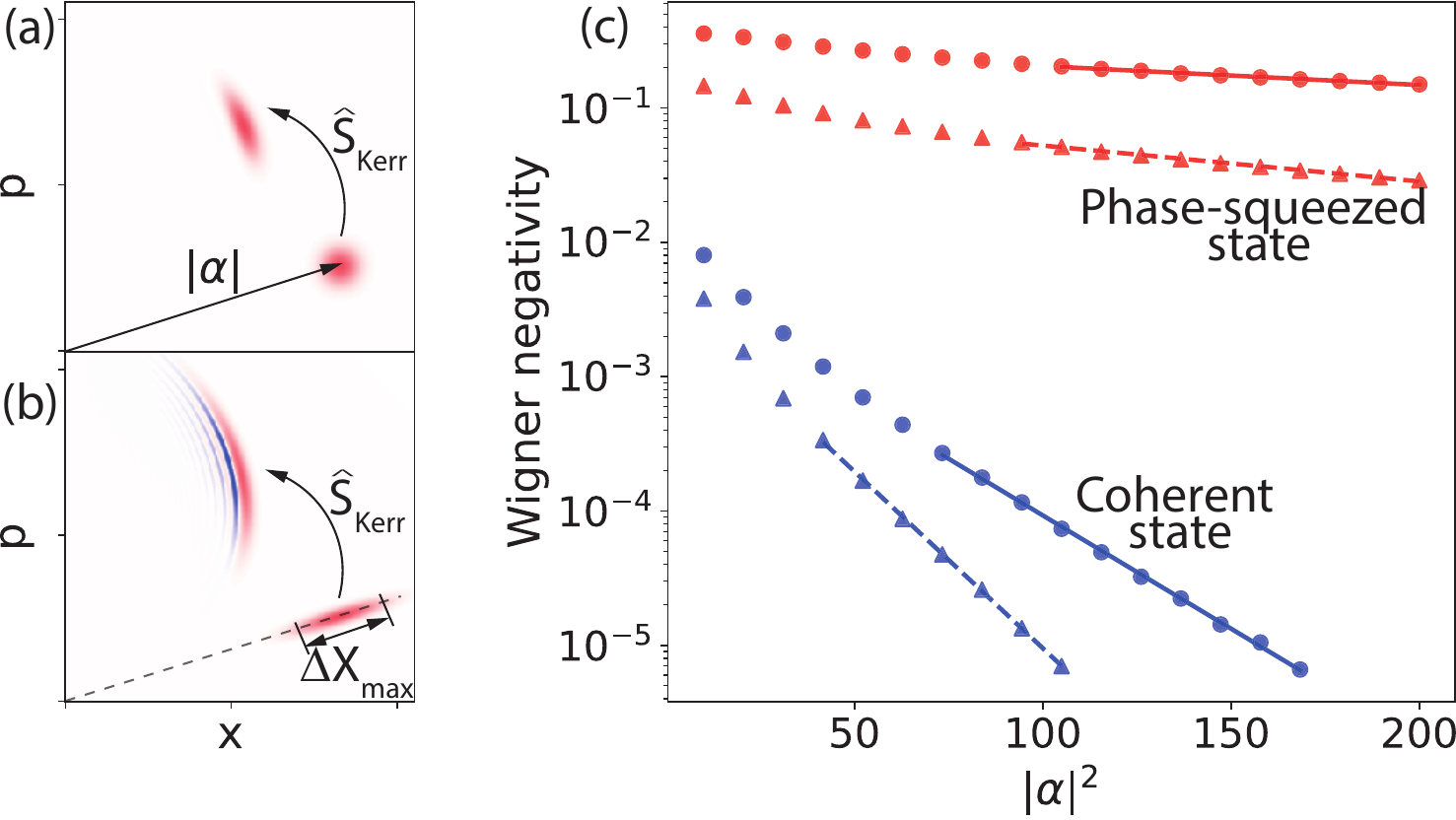}
     \caption{Transformation of the Wigner function due to the Kerr nonlinearity described by unitary transformation $\hat{S}_{\operatorname{Kerr}}$ for (a) a coherent state with $|\alpha|^2=200$ and (b) a phase-squeezed state with the same mean number of photons and degree of squeezing 8 dB. (c) Simulated Wigner negativity volume vs. the number of photons $|\alpha|^2$ for the fixed Kerr phase $\phi_{\operatorname{Kerr}} = 2 \chi t |\alpha|^2 = 0.6$. Circles correspond to the lossless case, triangles assume a 5\%-loss before the Kerr interaction and 5\% loss after it. Lines visualize an exponential fit of the points.}
     \label{fig:negativity}
 \end{figure}

As the mixed BSV state is composed of pure squeezed states stochastically displaced predominantly in the x-quadrature direction, one has to address the Kerr evolution of these displaced state. In the limit of weak Kerr nonlinearity, a coherent state of light experiences predominantly quadrature squeezing with an extremely low level of Wigner-function negativity. At the same time, some components of the statistical mixture in Eq. \eqref{eq:sq_th}, namely those resembling phase-squeezed states, are expected to have a very pronounced Wigner-function negativity under the same weak action of Kerr nonlinearity that leads to quadrature-squeezing of the coherent state. Figure~\ref{fig:negativity}(a,b) compares the simulated effect of the Kerr nonlinearity on the Wigner functions of a coherent state and a phase-squeezed state of the same energies (see Section~\ref{sec:kerr_simm} for the details). The considered level of squeezing is 8 dB, corresponding to $\operatorname{sinh}^2 r = 1$ photon in the squeezing operator. The accumulated nonlinear Kerr phase $\phi_{\operatorname{Kerr}}$ is assumed to be the same as in our experiment with BSV, $\phi_{\operatorname{Kerr}}=0.6$; however, the photon number of the simulated state is considerably lower ($|\alpha|^2 = 200$).

Figure~\ref{fig:negativity}(c) shows the dependence of the volume of the Wigner-function negativity of an initially coherent (blue points) and an initially phase-squeezed (red points) state on the initial displacement $|\alpha|$ for a fixed nonlinear phase $\phi_{\operatorname{Kerr}}=0.6$. In both cases, circles correspond to no losses and triangles, to 5\% losses before the Kerr effect and another 5\% losses after it.
As the displacement increases, the value of the Kerr nonlinearity is assumed to drop, to keep the nonlinear phase constant. This leads to an exponential reduction of the Wigner negativity with increasing photon number.
At the same time, even a moderate (8 dB) phase squeezing of the state amplifies the Wigner negativity by several orders of magnitude (Fig.~\ref{fig:negativity}(c)). 
Moreover, it improves the exponential decay rate of the Wigner negativity with the number of photons by more than a factor of 10. 
This allows us to expect substantial Wigner negativity for thousands of displacement photons for a phase-squeezed light, while in the case considered for the coherent pump, negativity gets vanishingly small for $|\alpha|^2 \sim 100$. 

The reason for this behavior is that a phase-squeezed state has reduced phase noise and amplified amplitude noise, which makes it more sensitive to the intensity-dependent Kerr phase.
Different parts of the Wigner function along the radius (dashed line in Fig.~\ref{fig:negativity}(b)) experience different nonlinear phases, resulting in a non-Gaussian shape and, as a consequence, in negativity. 
This effect allows for the observation of a  significant Wigner negativity for much brighter states even in the presence of losses. 

\section{Discussion}\label{sec:disc}
We have experimentally studied the effect of the Kerr nonlinearity on a macroscopic state of light, namely single-mode BSV. In order to sample the Husimi function, we used a single-shot f-2f interferometer, which provided the simultaneous measurement of the BSV amplitude and phase. At moderate Kerr interaction, the measured Husimi function acquired an S-shape, a direct consequence of the nonlinear intensity-dependent phase. Under these experimental conditions, the photon-number statistics of BSV was still unchanged, and the state remained single-mode (see SI, Section 1). At stronger Kerr interaction, which was achieved by focusing BSV tighter into the nonlinear material, the experimentally measured Husimi function deviated from the one expected from simulation using the classical Liouville equation. In addition, this regime led to the appearance of higher-order frequency modes and to a modified photon-number statistics (SI, Section 1). We assign all these effects to additional nonlinear optical processes, which appear at high intensity.

The non-Gaussian shape of the Husimi function is an evidence of the non-Gaussianity of the state. For a pure state, this would lead to the negativity of the Wigner function. However, the purity of a macroscopic state such as BSV is very susceptible to loss. Assuming a loss of 5\%, which can be caused by imperfect anti-reflection coating on the nonlinear crystal and lenses, imperfect reflectivity of the mirrors, as well as the nonlinear effects occurring in the nonlinear crystal, we can estimate the purity of our BSV as $\mathcal{P} \sim10^{-5}$. This low purity definitely prevents the direct observation of the Wigner-function negativity.

However, even after loss, BSV still contains some pure quantum superpositions. We have shown that a realistic BSV can be represented as a mixture of states squeezed along the same quadrature, which is close to the phase quadrature for states with large displacements. This answers the long-standing question \cite{Rasputnyi2024, Heimerl2025} to what extent can BSV be treated as a quantum superposition. 
To fully harness quantum features, such as Wigner function negativity, BSV must be distilled to isolate a pure squeezed coherent state.
Adopting the purification protocol for Gaussian states \cite{Glockl2006} and developing new protocols based on the strong-field light-matter interaction presents a clear avenue for future work, enabling exploration of quantum dynamics beyond the statistical limit.

\section{Methods}\label{sec:methods}

\subsection{Quantum and classical simulations of Kerr interaction}\label{sec:kerr_simm}
The optical Kerr effect arises from the third–order nonlinear susceptibility of a medium. In terms of quantum optics, for a single mode with bosonic annihilation operator $\hat{a}$, it can be described by the effective Hamiltonian
\begin{equation}
    \hat H_K =  \hbar \chi ~\hat a^\dagger{}^2 \hat a^2,
    \label{eqn:Kerr_hamiltonian}
\end{equation}
where $\chi$ (with units of angular frequency) is the Kerr interaction constant. For multi‑mode fields, cross‑Kerr interactions between different modes also arise, but here we focus on the single‑mode case and the self‑Kerr interaction effect. The corresponding unitary evolution operator for an interaction time $t$ is
\begin{equation}
    \hat{S}_{\operatorname{Kerr}}(t) = \exp\!\bigl(-i \chi t ~ \hat{a}^{\dagger 2}\hat{a}^{2}\bigr).
    \label{eqn:kerr_unitary}
\end{equation}
The transformation of the Wigner function under this process can be described with the following Wigner current in the polar coordinates $W(r,\phi)$\footnote{The chosen polar coordinates system is rotating with angular velocity $\chi$ to compensate for the rigid rotation of the Wigner function, caused by the chosen form of the Kerr Hamiltonian} \cite{Rosiek2024} :
\begin{equation}
    \vec{J} = \chi \Big(-2 r^2~ W + \frac{1}{8} \nabla^2 W \Big) r \vec \phi.
    \label{eqn:wigner_current}
\end{equation}
The first term, proportional to $r^2$, describes the azimuthal flow of the Wigner function, with the angular speed proportional to the squared distance to the origin. This corresponds to a photon‑number–dependent phase shift, which leads to the shearing of quantum states in phase space. The second term contains higher-order spatial derivatives of the Wigner function, and it is responsible for the appearance of the negative values of the Wigner function. The complete quantum simulation of the Kerr effect requires taking into account the complete dynamics, described by either the Wigner current in Eq. \eqref{eqn:wigner_current} or the unitary $\hat{S}_{\operatorname{Kerr}}$ in Eq. \eqref{eqn:kerr_unitary}. We perform the simulation shown in Fig.~\ref{fig:negativity} by representing the Kerr unitary operator $\hat{S}_{\operatorname{Kerr}}$ in the Fock basis and implementing the evolution using the Python toolbox QuTiP. This approach provides a flexible framework for analyzing the dynamics of the quantum state and its Wigner function, but the truncation to the lowest 700 Fock states we use here  restricts the simulations to relatively low‑energy states. The simulations clearly reveal narrow (so-called sub-Planckian) regions where the Wigner function of the Kerr‑evolved phase-squeezed states becomes negative. The overall degree of Wigner negativity is quantified by the negativity volume 
\begin{equation}
N_{\mathrm{neg}}[W] = - \int_{W(x,p)<0} \,W(x,p)\,dx\,dp,
\end{equation}
which is shown in Fig.~\ref{fig:negativity}(c). For completeness, we also show the Wigner function of the Kerr‑evolved squeezed vacuum state in the Supplementary Materials.

In our experiment, the noise of the Husimi function measurement is much larger than the vacuum-level quadrature fluctuations. Thus, any fine features of the quantum state cannot be directly observed there. In this regime, higher-order derivatives in the Wigner current \eqref{eqn:wigner_current} can be neglected to describe the measured coarse-grained Husimi functions of our high photon-number states. This brings the dynamics of the states to the classical Liouville equation, which includes only energy-dependent rotation of the phase space (the first term in Eq. \eqref{eqn:wigner_current}).

The value of the Kerr interaction constant $\chi$ is related to the non-linear refraction index of the medium $n_2$ as
\begin{equation}
    \chi = \frac{\hbar \omega_{0}^{2} c}{n_{0}^{2} V_{\mathrm{eff}}}\,n_{2},
\end{equation}
with $\omega_{0}$ the angular frequency of the light,  $n_0$ the refraction index, and $V_{\mathrm{eff}}$ the effective mode volume.
The non-linear refractive index $n_2$ used in the simulation of Fig.~\ref{fig:husimi} corresponds to values reported in the literature for fused silica, $n_2 \sim 10^{-20}$ m$^2$/W \cite{Schiek:23}.

\subsection{Quantum decoherence of BSV}\label{sec:mixed_BSV}

Optical linear losses are modeled as a beamsplitter, which reflects part of the radiation but adds a contribution of the vacuum from the `empty' input port.
As a result, the input variance $\Delta X_{in}^2$ of a quadrature becomes
\begin{equation}
    \Delta X_{out}^2 = T \Delta X_{in}^2 + R \Delta X_{vac}^2,
\end{equation}
where $R$ is the reflection coefficient of the beamsplitter, which is equivalent to losses, $T = 1-R$ is the transmission coefficient, and $\Delta X_{vac}^2=1$ is the vacuum quadrature.
As a result, the losses reduce the purity of BSV:
\begin{equation}
    \mathcal{P} = \dfrac{1}{\sqrt{\Delta X_{max}^2 \Delta X_{min}^2}} = \dfrac{1}{\sqrt{1 + 4RTN}} \approx \dfrac{1}{2\sqrt{RTN}},
\end{equation}
where $N = \sinh^2(r_0)$ is the mean number of photons of BSV and $r_0$ is the initial squeezing parameter.
This shows explicitly that one has to manage losses up to $R \sim 1 / N$ to have a pure quantum state of BSV.
After linear losses, BSV remains a Gaussian state of light, which can be described by squeezed $\Delta X_{min}^2$ and anti-squeezed $\Delta X_{max}^2$ quadratures.
By considering the two quadratures after losses, we can represent BSV as a squeezed thermal state, in which the initial mean number of  photons is $n_{th}$ and the squeezing parameter is $r$:
\begin{align}
    \Delta X_{min}^2 &= T e^{-2 r_0} + R = (1 + 2n_{th}) e^{-2r},\\
    \Delta X_{max}^2 &= T e^{2 r_0} + R = (1 + 2n_{th}) e^{2r}.
\end{align}
From these expressions we can estimate the squeezing parameter $r$ and the number of thermal photons $n_{th}$:
\begin{align}
    e^{2r} &= \sqrt{\dfrac{Te^{2r_0} + R}{Te^{-2r_0} + R}}\\
    n_{th} &= \dfrac{\sqrt{1 + 4 RT N} - 1}{2}
    \label{eq:th_sq}
\end{align}
In the limit of a large photon number $N$ and moderately low optical losses $R \sim 0.1$, we can simplify Eqs.~(\ref{eq:th_sq}):
\begin{align}
    r &\approx \dfrac{r_0}{2} - \dfrac{1}{4}\ln(R/T),\\
    n_{th} &\approx \sqrt{RT N}.
\end{align}

The resulting state of BSV after losses can be written as
\begin{equation}
    \hat{\rho}_{\text{BSV}} = \hat{S}(r) \hat{\rho}_{th} \hat{S}^{\dagger}(r) = \int d^2\alpha~ P(\alpha) \hat{S}(r) |\alpha\rangle \langle \alpha| \hat{S}^{\dagger}(r),     
    \label{eq:rho_BSV}
\end{equation}
where 
\begin{equation}
    P(\alpha) = \dfrac{1}{\pi n_{th}} \exp\bigg(-\dfrac{|\alpha|^2}{n_{th}}\bigg)
    \end{equation}
is the Glauber–Sudarshan P representation for a thermal state. The density matrix in Eq.~(\ref{eq:rho_BSV}) can be rewritten as
\begin{equation}
    \hat{\rho}_{\text{BSV}}=\int d^2\beta~ \widetilde{P}(\beta) |\beta, r\rangle \langle \beta, r|,
\end{equation}
which means that BSV after losses is a mixture of squeezed coherent states $|\beta, r\rangle$, with weights forming a double-Gaussian distribution: 
\begin{equation}
    \widetilde{P}(\beta) = \dfrac{1}{\pi n_{th}} \exp\left[ -\dfrac{\text{Re}(\beta)^2}{n_{th} e^{2r}}-\dfrac{\text{Im}(\beta)^2}{n_{th} e^{-2r}} \right].
\end{equation}

Although the density matrix of the lossy BSV in equation~(\ref{eq:rho_BSV}) is unique, its decomposition into pure states is mathematically non-unique. For instance, the thermal state can equivalently be decomposed in the photon-number Fock basis, treating BSV as a mixture of squeezed Fock states. Consequently, the fundamental nature of a single BSV pulse realization cannot be uniquely determined independently of the measurement scheme.

It is also worth noting that the impurity of the BSV state after losses arises due to entanglement with the environment, following the quantum beamsplitter model. Tracing out the environment results in a squeezed thermal state where individual pulses do not strictly possess a pure wavefunction. Physically, this mechanism is distinct from a `proper mixture' where a pure squeezed coherent state is subject to random displacements. However, in the limit of local measurements where the environment is inaccessible, these two mechanisms of decoherence are operationally indistinguishable \cite{Zurek2003}. This equivalence justifies modeling the macroscopic Kerr effect using an ensemble of pure squeezed coherent state trajectories.

\backmatter



\bmhead{Acknowledgments}
A.R. thanks Maksim Sirotin for helpful discussions.
This research was supported by the Deutsche Forschungsgemeinschaft (DFG, German Research Foundation), Project-ID 545591821. D.V.S. acknowledges his work to have been partially funded by the European Union's Horizon Europe Research and Innovation Programme under agreement 101070700 (project MIRAQLS).



\bibliography{bibliography}

\end{document}


\title[High Harmonic Generation by Bright Squeezed Vacuum]{High Harmonic Generation by Bright Squeezed Vacuum}


\author[1,2]{\fnm{Andrei} \sur{Rasputnyi}}

\author[3]{\fnm{Ilya} \sur{Karuseichyk}}

\author[1]{\fnm{Gerd} \sur{Leuchs}}

\author[1,4]{\fnm{Francesco} \sur{Tani}}

\author[5]{\fnm{Denis} \sur{Seletskiy}}

\author*[1,2,6]{\fnm{Maria} \sur{Chekhova}}\email{maria.chekhova@mpl.mpg.de}

\affil[1]{Max Planck Institute for the Science of Light, Erlangen, Germany,}

\affil[2]{Department of Physics, Friedrich-Alexander-Universität Erlangen-Nürnberg (FAU), Erlangen, Germany}

\affil[3]{Laboratoire d’Optique Appliqu\'ee (LOA), CNRS, \'Ecole polytechnique, ENSTA, Institut Polytechnique de Paris, Palaiseau, France}

\affil[4]{University of Lille, CNRS, UMR 8523—PhLAM—Physique des Lasers Atomes et Molécules, Lille, France}

\affil[5]{femtoQ Laboratory, Department of Engineering Physics, Polytechnique Montréal, 2500 ch. de Polytechnique, Montréal, QC H3T 1J4, Canada}

\affil[6]{Faculty of Electrical and Computer Engineering, Technion—Israel Institute of
Technology, Haifa 32000, Israel}



\section{Spectral intensity correlations and photon statistics}\label{sec:spectrum}

 \begin{figure}[h!]
     \centering
     \includegraphics[width=1.00\linewidth]{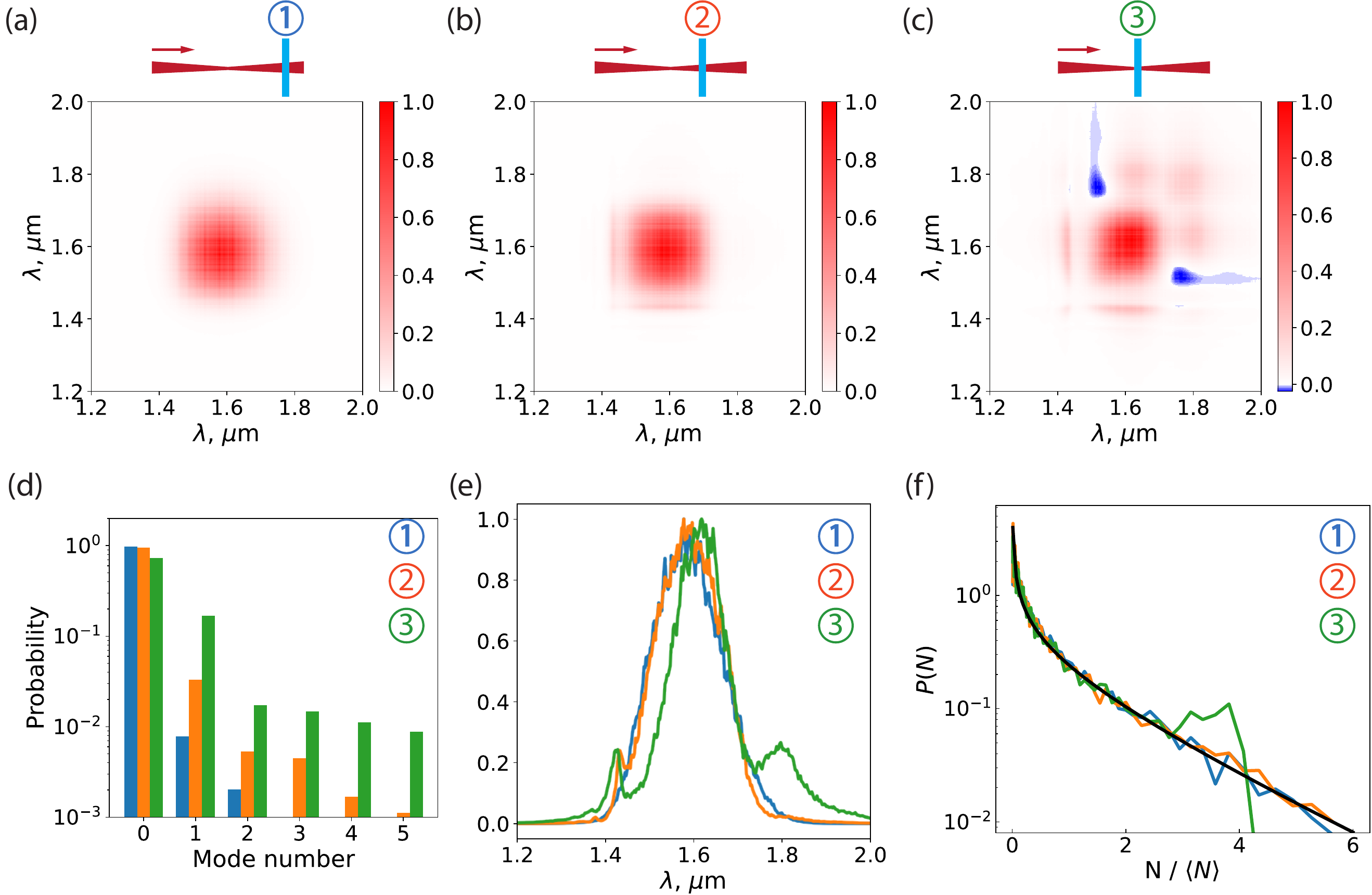}
     \caption{(a-c) Spectral intensity covariance maps for different positions of fused silica slab. (d) Probability of occurrence of different spectral modes. (e) The spectral intensity of the fundamental frequency mode. (f) Photon-number probability distribution. Black curve is fit with the Gamma distribution.}
     \label{fig:stat}
 \end{figure}
We investigate how the Kerr effect modifies the spectral intensity correlations of BSV. 
To this end, we measure the spectral intensity covariance maps, defined as $\text{Cov}(\lambda_1, \lambda_2) = \langle I(\lambda_1)I(\lambda_2) \rangle - \langle I(\lambda_1)\rangle \langle I(\lambda_2) \rangle$, where $\langle ...\rangle$ is averaging over the acquired BSV pulses. 
The spectral intensity covariance is measured by acquiring 7000 single-shot spectra of BSV.
Figures~\ref{fig:stat}(a-c) show the covariance maps obtained for different positions of the fused silica slab, corresponding to different magnitudes of the Kerr interaction.
We see that while the covariance has a  close-to-Gaussian shape in the case of weak Kerr effect (Fig.~\ref{fig:stat}(a)), stronger Kerr interaction leads to the appearance of additional lobes and even negativity (Fig.~\ref{fig:stat}(c)). 
The negativity (not to be confused with the Wigner-function negativity) indicates anti-correlation between different spectral intensity components, which by itself is not a quantum effect.

By performing the singular value decomposition of the covariance matrix, we can estimate the spectral mode content of the radiation: the weights of the frequency modes (Fig.~\ref{fig:stat}(d)) and their spectral shapes. As an example, Fig.~\ref{fig:stat}(e) shows the shape of the fundamental mode.
By integrating the spectral intensity of each shot, we can also characterize the photon-number probability distribution (Fig.~\ref{fig:stat}(e)).
The photon-number distribution of single-mode BSV follows the Gamma distribution $P(N) = \dfrac{1}{\sqrt{2\pi N \langle N \rangle}} \exp\bigg(-\dfrac{N}{2\langle N \rangle}\bigg)$, where $\langle N \rangle$ is the mean photon number, which is plotted as a black line in Figure~\ref{fig:stat}(f).
In this experiment, the BSV pulse energy is $130$ nJ, which corresponds to $\langle N \rangle \approx 1 \times 10^{12}$.

When the fused silica slab is placed far from the focus, a separable 2D-Gaussian distribution of intensity covariance (Fig.~\ref{fig:stat}(a)) is a clear indication that the BSV is single-mode.
Indeed, the mode weights (panel d) show that the fundamental mode is strongly dominant, with the second mode's  weight being two orders of magnitude lower. Moreover, the fundamental mode has a nearly Gaussian shape (Fig.~\ref{fig:stat}(e)), and the photon-number distribution (Fig.~\ref{fig:stat}(f)) follows the Gamma distribution, expected for single-mode BSV.
When the slab is in the intermediate position, we start to see slight spectral broadening (Fig.~\ref{fig:stat}(b)), an increase in the weights of the higher-order frequency modes (Fig.~\ref{fig:stat}(d)), but still a Gaussian fundamental mode and a photon-number distribution close to a single-mode one (Fig.~\ref{fig:stat}(e,f), orange curves).

If the silica window is placed in the focus, we observe significant spectral broadening (Fig.~\ref{fig:stat}(c)). Along with the continued increase in higher-order mode weights  (Fig.~\ref{fig:stat}(d)), the spectral shape of the fundamental mode also changes  (Fig.~\ref{fig:stat}(e)).
In this case, the tail of the photon-number distribution does not follow the exponential decay but has a sharp cut-off.
From a quantum-optical point of view, Kerr nonlinearity contributes only to the additional nonlinear phase without changing the photon-number probability distribution.
The observed modification of photon-number statistics is related to the additional nonlinear-optical processes accompanying supercontinuum generation.
These observations give us indirect evidence of the Kerr nonlinearity.

\section{Wigner function of a pure squeezed vacuum after Kerr interaction}\label{sec:pure_BSV}

As discussed in Ref.~\cite{Rosiek2024}, a pure squeezed vacuum state evolving under a Kerr nonlinearity develops pronounced Wigner negativity. Because the squeezed vacuum coherently includes components with drastically different field amplitudes, the amplitude‑dependent Kerr phase induces a strong shearing of the state in phase space, which in turn generates interference fringes with significant negative regions in the Wigner function. 

\begin{figure}[t]
     \centering
     \includegraphics[width=1.0\linewidth]{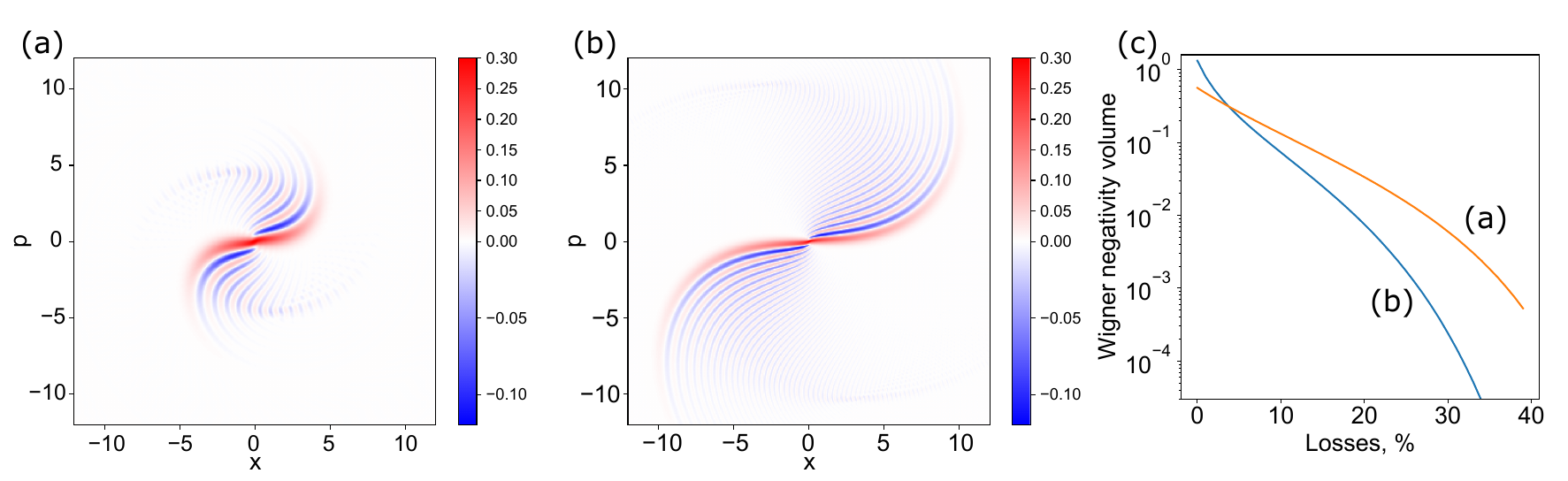}
    \caption{Wigner functions of Kerr-evolved squeezed vacuum states and their negativity under loss. The Kerr phase at the mean photon number $n_s$ is $\phi_{\mathrm{Kerr}} = 2\chi t n_s = 0.16$. (a) $n_s = 4$ (low energy, strong nonlinearity $\chi t = 0.02$). (b) $n_s = 20$ (higher energy, weaker nonlinearity $\chi t = 0.004$). (c) Wigner negativity volume versus loss after the Kerr interaction: orange curve for case (a), blue curve for case (b).}
     \label{fig:Wigner_SV_Kerr}
 \end{figure}

Two examples of Kerr-evolved squeezed vacuum Wigner functions are shown in Fig.~\ref{fig:Wigner_SV_Kerr}(a,b). These were obtained from Fock-basis simulations using the QuTiP Python toolbox, with initial mean photon numbers $n_s=4$ (a) and $n_s=20$ (b), both at fixed Kerr phase $\phi_{\mathrm{Kerr}} = 2\chi t n_s = 0.16$. The Kerr interaction transforms the initial elongated elliptical structure into an 'S'-shaped pattern, while generating numerous narrow interference fringes around the central maximum with alternating positive and negative regions. The fringe scale decreases as the nonlinearity weakens. Note that finer fringes of the Wigner function are more susceptible to loss, which acts as a Gaussian smoothing filter. This is illustrated in Fig.~\ref{fig:Wigner_SV_Kerr}(c), where the negativity volume $N_{\mathrm{neg}}$ of the brighter state with lower non-linearity (b) decays faster than that of the dimmer state with higher non-linearity (a), despite its initially larger negativity. While any pure non-Gaussian state retains some Wigner negativity up to $50\%$ loss \cite{nugmanov2022robustness}, moderate losses suffice to render bright-state negativity practically undetectable. 

\bibliography{bibliography}